\documentclass[reprint,amsmath,amssymb,aps,prl,superscriptaddress]{revtex4-1}

\usepackage{graphicx}
\usepackage{dcolumn}
\usepackage{bm}
\usepackage{hyperref}
\usepackage{comment}
\usepackage{siunitx}
\usepackage{multirow}

\usepackage{xcolor}

\usepackage{subcaption}
\makeatother

\newcommand{\betan}{$\beta_{N}$}
\newcommand{\fGW}{$f_{GW}$}
\newcommand{\qnf}{$q_{95}$}

\newcommand{\nebar}{$<$$n_e$$>$}
\newcommand{\densunit}{$\times$10$^{19}$ m$^{-3}$}
\newcommand{\In}{$I_P/(a B_T)$}
\newcommand{\Ip}{$I_P$}
\newcommand{\Bt}{$B_T$}

\newcommand{\Ptot}{$P_{tot}$}
\newcommand{\Pech}{$P_{ECH}$}

\newcommand{\IaB}{$I_p a B_t$}

\newcommand{\HH}{$H_{98y2}$}
\newcommand{\HL}{$H_{89}$}

\newcommand{\taue}{$\tau_E$}
\newcommand{\taueth}{$\tau_\text{E,th}$}

\newcommand{\Zeff}{$Z_{eff}$}

\begin{document}

\title{Simultaneous access to high normalized current, pressure, density, and confinement in strongly-shaped diverted negative triangularity plasmas}

\author{C. Paz-Soldan}
\affiliation{Columbia University, New York, NY 10027, USA}


\author{C. Chrystal}
\affiliation{General Atomics, San Diego, CA 92121, USA}

\author{P. Lunia}
\affiliation{Columbia University, New York, NY 10027, USA}

\author{A.O. Nelson}
\affiliation{Columbia University, New York, NY 10027, USA}


\author{K.E. Thome}
\affiliation{General Atomics, San Diego, CA 92121, USA}

\author{M.E. Austin}
\affiliation{University of Texas at Austin, Austin, TX 78712, USA}

\author{T.B. Cote}
\affiliation{General Atomics, San Diego, CA 92121, USA}

\author{A.W. Hyatt}
\affiliation{General Atomics, San Diego, CA 92121, USA}

\author{A. Marinoni}
\affiliation{MIT PSFC, Cambridge, MA, 02139 USA}

\author{T.H. Osborne}
\affiliation{General Atomics, San Diego, CA 92121, USA}

\author{M. Pharr}
\affiliation{Columbia University, New York, NY 10027, USA}

\author{O. Sauter}
\affiliation{Ecole Polytechnique Federale de Lausanne (EPFL), Swiss Plasma Center, Lausanne, Switzerland}

\author{F. Scotti}
\affiliation{Lawrence Livermore National Laboratory, Livermore, CA, 94550 USA}

\author{T.M. Wilks}
\affiliation{MIT PSFC, Cambridge, MA, 02139 USA}

\author{H.S. Wilson}
\affiliation{Columbia University, New York, NY 10027, USA}

\date{\today}

\begin{abstract}

Strongly-shaped diverted negative triangularity (NT) plasmas in the DIII-D tokamak demonstrate simultaneous access to high normalized current, pressure, density, and confinement. NT plasmas are shown to exist across an expansive parameter space compatible with high fusion power production, revealing surprisingly good core stability properties that compare favorably to conventional positive triangularity plasmas in DIII-D. Non-dimensionalizing the operating space, edge safety factors below 3, normalized betas above 3, Greenwald density fractions above 1, and high-confinement mode (H-mode) confinement qualities above 1 are simultaneously observed, all with a robustly stable edge free from deleterious edge-localized mode instabilities. Scaling of the confinement time with engineering parameters reveals at least a linear dependence on plasma current although with significant power degradation, both in excess of expected H-mode scalings. These results increase confidence that NT plasmas are a viable approach to realize fusion power and open directions for future detailed study.

\end{abstract}

\maketitle


\paragraph{Introduction:}

Tokamak plasmas featuring negative triangularity (NT) shaping have been proposed as a distinct path to solve the grand challenge of integrating a high-performance plasma core with a power handling solution at the edge \cite{Kikuchi2019,Marinoni2021} - a necessary step on the path to harnessing a burning plasma for fusion energy production. The power handling advantage is primarily due to the abandonment of the H-mode regime, which allows NT plasmas to operate with lower exhausted power crossing the separatrix (no longer constrained by staying above the power required for the H-mode transition), while also robustly avoiding the edge localized mode (ELM) instability \cite{Bishop1986,Saarelma2021,Nelson2022}. Furthermore, the NT configuration benefits from inherently larger surface areas for power exhaust due to the larger major radius of the exhaust region. These edge advantages are expected to come alongside improved core confinement arising from the direct effect of NT shaping on plasma microturbulence \cite{Pochelon1999,Camenen2007,Austin2019,Fontana2020,Coda2022,Happel2023}.

These NT benefits are usually contrasted with expected drawbacks in NT plasma stability \cite{Medvedev2008,Medvedev2015}. The implication is that the operating space in terms of important parameters such as plasma current, pressure, and density will necessarily be reduced as compared to conventional positive triangularity (PT) plasmas. Primarily for this reason, tokamak development has thus far focused on PT configurations, which are the basis for the design of the two next-step tokamaks designed to produce net fusion power gain: ITER \cite{Aymar2001} and SPARC \cite{Creely2020}.

\begin{figure}
    \includegraphics[width=.5\linewidth]{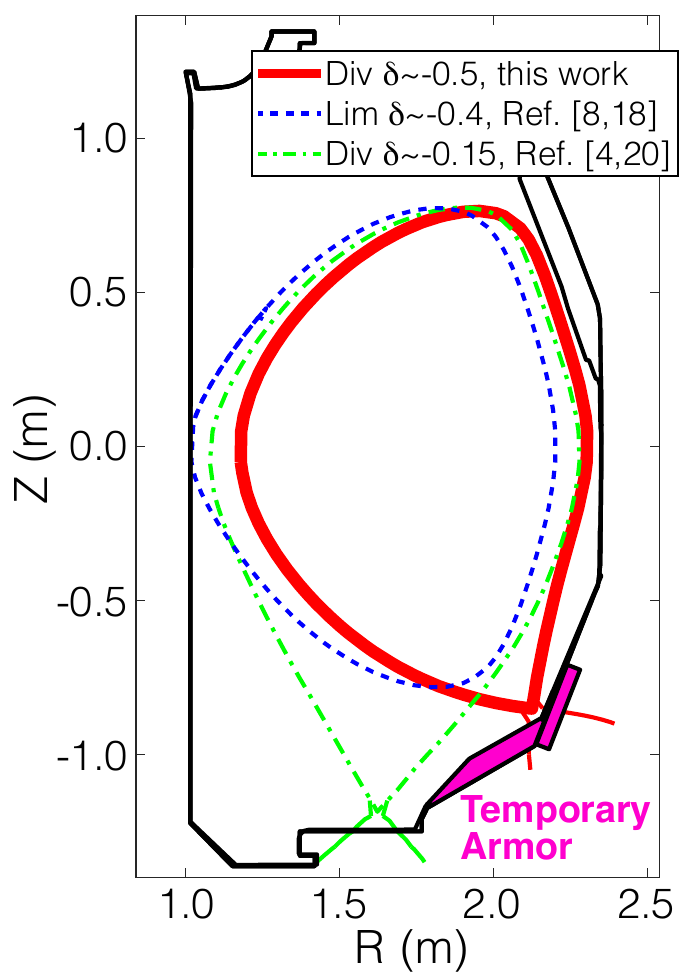}
    \vspace{-10 pt}
    \caption{NT shapes explored in DIII-D, with data in this Letter exclusively from the diverted strongly-shaped ($\delta \approx -0.5$) configuration, enabled by the installation of temporary power-handling structures (`armor').}
    \label{fig:shape}
    \vspace{-20 pt}
\end{figure}

\begin{figure*}[!ht]
    \includegraphics[width=0.8\linewidth]{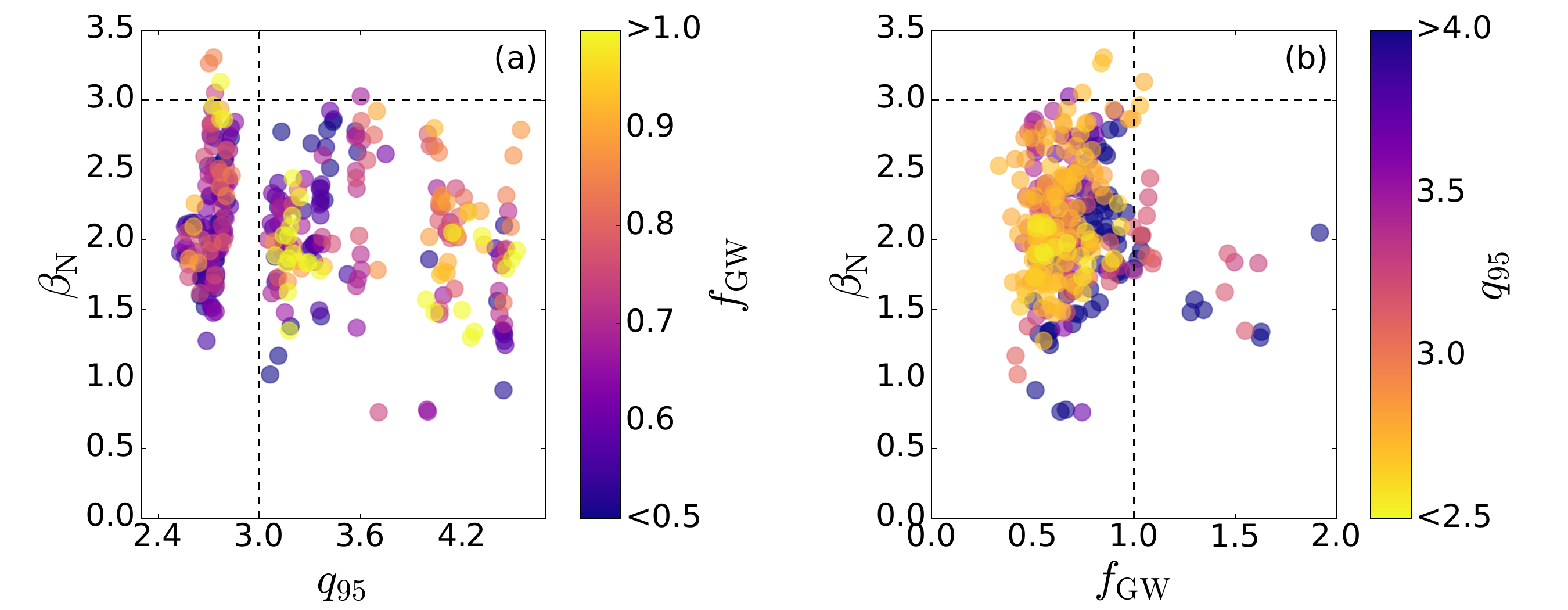}
        \vspace{-10 pt}
    \caption{Observed operating space for strongly-shaped ($\delta \approx -0.5$) diverted NT plasmas in terms of \qnf{}, \betan{}, and \fGW{}. Each datapoint is a $\Delta t \ge$ 3 \taue{} stationary phase. Reference values of \qnf{} = 3, \betan{} = 3 and \fGW{} = 1 are indicated.}
    \label{fig:opSpace}
    \vspace{-15 pt}
\end{figure*}

In this Letter, the operating space of strongly-shaped diverted NT plasmas in DIII-D is shown to be expansive, highly-performing, and in multiple respects improved over PT plasmas. This indicates overall that the perceived concerns with NT plasma stability are significantly less severe than previously expected, increasing confidence that NT plasmas are a viable approach to realize fusion conditions in the tokamak. High normalized plasma current, pressure, density are achieved simultaneously, yielding operating points outside of the boundaries conventionally assumed realistic for NT and even PT plasmas.  Furthermore, these operating points are shown to be accessed with high confinement by comparing to expectations from standard H-mode scaling laws \cite{IPBch2}. Extrapolation is briefly considered using basic engineering parameter regression of the measured confinement time (\taue{}, the ratio of plasma energy to input power), drawing comparisons to the established H-mode scaling \cite{IPBch2}. This work significantly advances the results of Ref. \cite{Austin2019} by presenting expanded operation in current and density while retaining the reported high pressure \cite{Boyes2023} and confinement \cite{Marinoni2019}. NT discharges also now integrate the edge topology of a poloidal divertor, as opposed to Ref. \cite{Austin2019}'s limited configuration (also shown in Fig. \ref{fig:shape}). A poloidal divertor is considered necessary for power and particle handling in tokamak-based fusion devices \cite{Leonard2018}.

\paragraph{Background:}

All experimental data presented in this Letter arise from unique strongly-shaped (triangularity $\delta \approx$ -0.5)  NT plasmas with a poloidal divertor, as shown in Fig. \ref{fig:shape}. These plasmas differ from past NT configurations in DIII-D, which were either strongly-shaped but with a limiter configuration \cite{Austin2019,Marinoni2019}, or with a divertor configuration but weakly-shaped \cite{Saarelma2021, Marinoni2021a}. Strongly-shaped diverted NT plasmas are enabled by the installation of temporary power-handling graphite components (`armor'), covering penetrations for diagnostic access. Using this armor, hundreds of NT discharges were produced during a dedicated experimental campaign, with the results in this paper spanning plasma current \Ip{} = 0.5-1.1 MA, toroidal field \Bt{} = 1.0-2.1 T, line-averaged density \nebar{} = 3-12\densunit{}, and input power \Ptot{} = 0.5-12 MW. \Ptot{} consisted of Ohmic, neutral beam, and electron cyclotron heating (\Pech{}). All injected power is taken to be absorbed by the plasma. A key goal of the campaign is the identification of operational limits in plasma current, pressure, and density, as well as documentation of the confinement quality observed as these parameters vary.

The operational limits of both PT and NT tokamaks are thought to be governed by the same fundamental physical mechanisms. The current limit is due to the external kink instability, modified by resistivity \cite{Wesson1978,Turnbull2016}. The relevant figure of merit is the edge safety factor \qnf{}, whose variation from PT to NT is a straight-forward calculation \cite{Sauter2016}. For both PT and NT, a hard limit exists at \qnf{} = 2, and a soft limit exists at \qnf{} = 3 where instability is often found \cite{Wesson1989}. The preponderance of existing PT reactor designs limit themselves to \qnf{} $\gtrsim$ 3 for their operational point, despite the fact that higher confinement is projected below this value \cite{Aymar2001}. The pressure limit is also due to the external kink, though driven by pressure as well as current \cite{Troyon1984}, with resistive tearing instabilities found approaching the ideal limit \cite{Brennan2003}. The relevant figure of merit is the normalized beta \betan{} \cite{Troyon1984}. Modeling generically finds reduced ideal \betan{} limits for NT plasmas as compared to PT plasmas \cite{Medvedev2008,Medvedev2015}.The density limit is still incompletely understood \cite{Greenwald2002}, with mechanisms focusing on the implications of density-driven edge cooling to resistive stability, edge radiative collapse and/or shear flow collapse. Candidate resistive instabilities include core tearing modes \cite{Gates2012} and interchange modes in the scrape-off-layer \cite{Giacomin2022}. These mechanisms do not depend directly on triangularity. The relevant empirical metric for density access is the Greenwald fraction \fGW{} ($\equiv{}$ \nebar$\pi a^2/I_p$, where $a$ is the minor radius) \cite{Greenwald1988}.

\begin{figure}[b]
        \vspace{-25 pt}
    \includegraphics[width=0.9\linewidth]{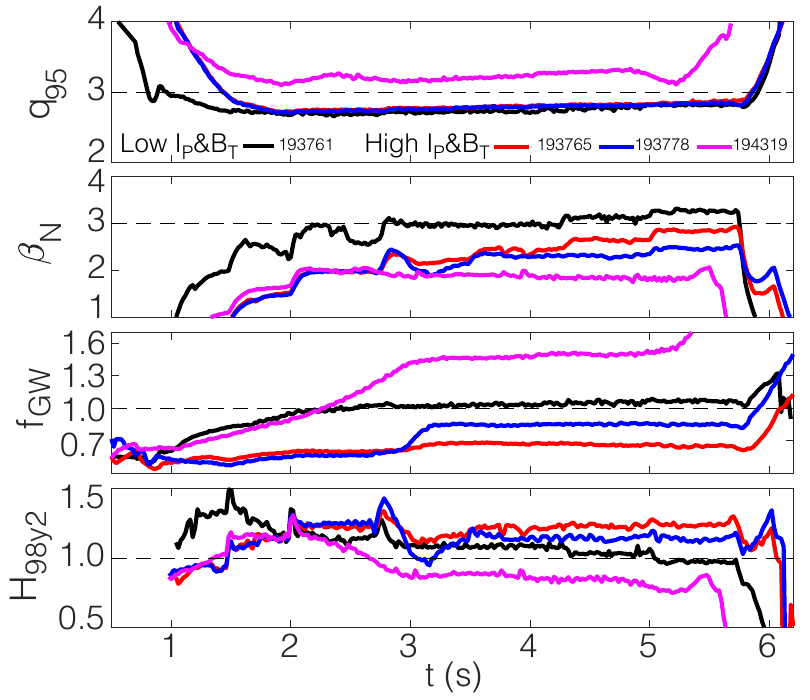}
        \vspace{-10 pt}
    \caption{Strongly shaped-diverted NT discharges illustrating long-duration simultaneous access to high performance in terms of \qnf{}, \betan{}, \fGW{}, and \HH{}.}
    \label{fig:time}
    \vspace{-15 pt}
\end{figure}

\begin{figure*}[!ht]
    \includegraphics[width=0.95\linewidth]{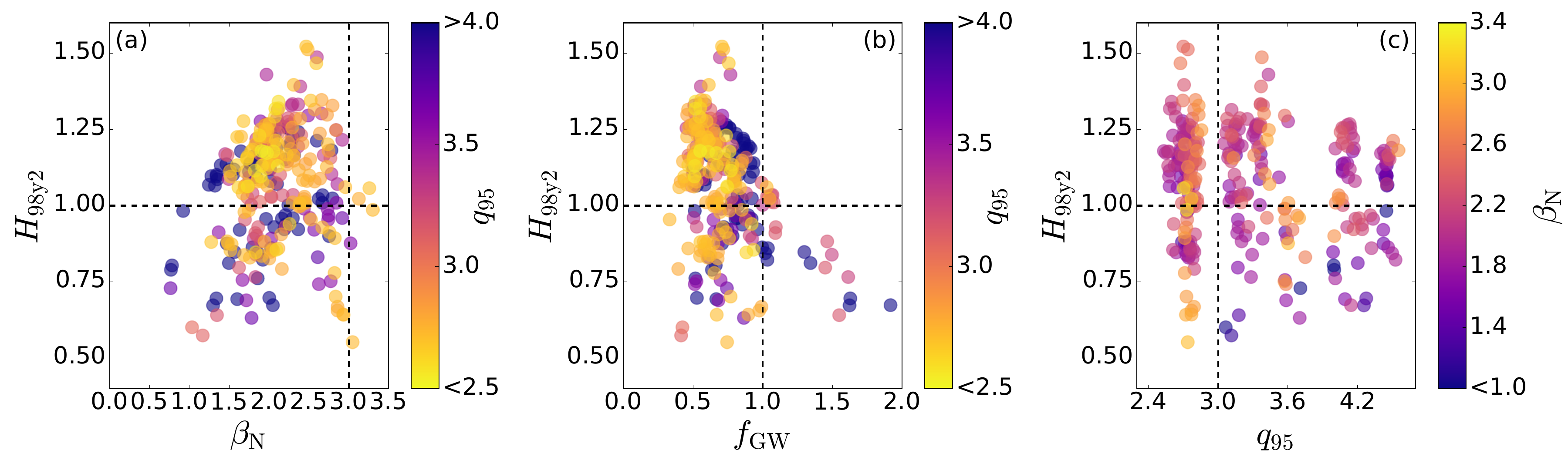}
        \vspace{-10 pt}
    \caption{Observed confinement quality (\HH{}) for strongly-shaped ($\delta \approx -0.5$) diverted NT plasmas in terms of \betan{}, \fGW{}, and \qnf{}. Each datapoint is a $\Delta t \ge$ 3 \taue{} stationary phase. \HH{} $>$ 1.4 is only found in low \Ip{} \& \Bt{} discharges.}
    \label{fig:Hfactor}
        \vspace{-15 pt}
\end{figure*}


\paragraph{Operating Space:} 

A comprehensive database of the strongly-shaped diverted NT plasma operating space is prepared by filtering the complete strongly-shaped ($\delta \approx -0.5$) dataset to remove cases with impurity seeding \cite{Kallenbach2013} and with saturated tearing mode instabilities \cite{Buttery2000a}. Remaining cases exhibit small and rapid core sawtooth instabilities \cite{Hastie1997}. Stationary phases of $\Delta t \gtrsim 400$ ms long ($\approx$ 3 \taue{}) were identified, imposing that each have a normalized standard deviation $\sigma_\mathrm{x}/\overline{x}<3$\% and a normalized average gradient $(dx/dt) \times (\Delta t/\overline{x})<0.25$\%, where $x$ included \betan{}, $W_\mathrm{MHD}$, \qnf{} and \nebar{}. Note effective charge (\Zeff{}) $<$ 2 is predominant across the dataset despite significant armor tile misalignments ($\lesssim$ 1 cm).

As shown in Fig. \ref{fig:opSpace} an exceptionally wide operating space is found, demonstrating access to low \qnf{}, high \betan{}, and high \fGW{} (and all without ELMs). Indeed, along some dimensions such as \qnf{} and \betan{}, the NT operating space clearly exceeds the expectation of PT plasmas. Operation at low \qnf{} is thought to be primarily facilitated by the absence of ELMs, which in PT become very large and pernicious in the \qnf{} $<$ 3 regime \cite{Hanson2014,Merle2017,Sauter2020}. PT L-modes can operate at low \qnf{} \cite{Piovesan2014,Hanson2014}, but accessing high \betan{} necessitates entering H-mode. In PT H-modes at \qnf{} $\lesssim$ 3, \betan{} $\gtrsim$ 3 is near observed resistive instability limits \cite{Sauter2002,Sauter2002a}. The underlying reason for this will be the subject of detailed analysis, but the most compelling hypothesis relates to the complete absence of ELMs alongside an observed weaker sawtooth yielding a much more benign landscape in terms of MHD seeding events. Sawteeth are observed to be weaker in NT \cite{Reimerdes2000}, supported by calculations that find increased sawtooth growth rates in NT vs PT \cite{Martynov2005,Liu2022a} which counter-intuitively manifests as higher frequency smaller crashes. The pressure and current gradients are also greatly modified, possibly improving the classical resistive stability systematically.

Operation at high \fGW{} is also remarkable, with PT reactor designs assuming \fGW $<$ 1, despite fusion reactivity clearly favoring higher density. Note the presented \fGW{} corresponds to the calculation from the line averaged \nebar{}. Due to the strongly peaked nature of NT density profiles \cite{Austin2019}, high \nebar{} and \fGW{} are achieved in this dataset without ever violating the empirical density limit (local \fGW{}$>1$) at the edge. The ability to operate with \fGW{} $\gg$ 1 is thus an observed generic feature of NT plasmas in DIII-D, with detailed transport calculations required to extrapolate this finding to fusion conditions.


To further highlight the database findings of Fig. \ref{fig:opSpace}, long-duration stationary discharges accessing high normalized parameters are shown in Fig. \ref{fig:time}. Parameter access is demonstrated simultaneously, such as access to high \betan{} and \fGW{} alongside very low \qnf{} $<$ 3. Confinement quality (\HH) is presented as compared to the expectation of the established H-mode scaling law \cite{IPBch2}, where \HH{} $>$ 1 is considered better than standard H-mode confinement. Degradation of \HH{} is found as \betan{} $\gtrsim$ 3, and severe degradation of \HH{} is found as \fGW{} $\gtrsim$ 1, correlated with increased radiation. Also worth noting that some of the discharges with the highest \HH{} occurred at low-field (\Ip{}/\Bt{} = 0.6 MA / 1.2 T) rather than full-field (\Ip{}/\Bt{} = 1.0 MA / 2.0 T) conditions. 



\paragraph{Confinement Quality:}

Attention now turns to a statistical view of the confinement quality observed across the NT operational space, shown in Fig. \ref{fig:Hfactor}. Some \HH{} values are exceptionally high $>$ 1.4, a regime exclusively found at low \Ip{} \& \Bt{}. Even excluding the cases with \HH{} $>$ 1.4, optima in \HH{} are found against several parameters. A clear optimum is found at \betan{} $\approx$ 2.5, with lower \HH{} values found above and below, well away from the maximum achieved \betan{} $\gtrsim$ 3. The trend of increasing maximal \HH{} with \betan{} is consistent with either an increased critical gradient or reduced stiffness of turbulent transport, with both mechanisms requiring high power operation to manifest \cite{Mantica2011,Hillesheim2013}.

A similarly clear optimum in \HH{} is found around \fGW{} $\approx$ 0.7. Above this value a striking decrease in peak \HH{} is found through to the operational limit at \fGW{} $\approx$ 2. This is consistent with the absolute \taue{} remaining roughly invariant with \nebar{}, possibly due to core turbulence changes reminiscent of the Linear to Saturated Ohmic Confinement (LOC/SOC) regime change \cite{Rice2013}, or due to an increase of turbulence in the SOL and across the separatrix \cite{Sauter2014,Giacomin2022}.

Finally, no clear \HH{} optimum with \qnf{} is found, suggesting the predicted \HH{} scaling for \Ip{} is valid. The maximal \HH{} instead is found to continually rise with reduced \qnf{} despite sawtoothing being present. This finding is contrary to work in PT, where a reduction in \HH{} was found in H-mode plasmas for \qnf{} $\lesssim$ 3.5 \cite{Schissel1992}. In PT, the degraded confinement was interpreted to be due to a significantly enlarged and impactful sawtooth. As with the improved high \betan{} and low \qnf{} operational range shown in Fig. \ref{fig:opSpace}(a), the improved confinement at low \qnf{} is proposed to be due to the relatively benign sawtooth in NT \cite{Martynov2005,Liu2022a} as compared to PT, alongside the complete absence of ELMs \cite{Nelson2023a}.


\begin{figure}
    \includegraphics[width=1\linewidth]{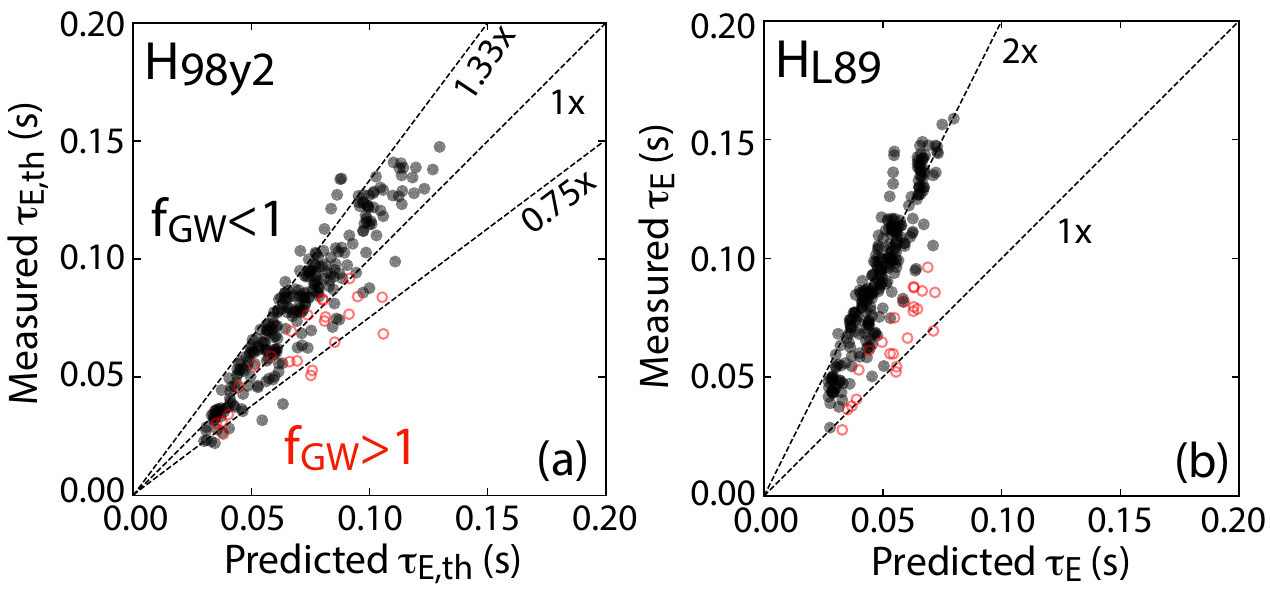}
        \vspace{-20 pt}
    \caption{Comparison of \taueth{} and \taue{} to scaling law expectation for (a) H-mode (\HH{}) and (b) L-mode (\HL{}). Data with \fGW{} $>1$ is differentiated as it falls below the \fGW{} $<1$ trend.}
    \label{fig:regress}
    \vspace{-15 pt}
\end{figure}

\paragraph{Confinement Scaling via Engineering Regression:} 

Consistency of the discussed NT database with the expectations of the \HH{} and the L-mode scaling law (\HL{} \cite{Yushmanov1990}) is presented in Fig. \ref{fig:regress}(a,b). Considering the \HH{} scaling in Fig. \ref{fig:regress}(a), though a range of \HH{} values were shown in Fig. \ref{fig:Hfactor}, an overall good agreement with the \HH{} prediction is still found. Interestingly, the measured \taueth{} is systematically below (above) the scaling law expectation at low (high) \taueth{}. Note comparing confinement to the H-mode scaling requires subtracting the fast particle contribution to \taue{}, yielding \taueth{}. This is here done via the standard DIII-D between-shot analysis routines \cite{Heidbrink1994}. Similarly, considering \HL{} in Fig. \ref{fig:regress}(b), a more positive slope is seen. \HL{} = 1 is found only at low \taue{}, while at high \taue{}, \HL{} $\approx$ 2 is approached. Due to these clear differences in slope, new analysis for the confinement scaling of NT plasmas is warranted.

As is typical, a regression for \taueth{} can be made with basic engineering parameters: \Ip{}, \Bt{}, \nebar{}, and \Ptot{} \cite{Kaye1985,Christiansen1992}. Since the database contains effectively no variation in plasma size, shape, or isotope mass, these parameters are not considered. To avoid sampling bias, weights are assigned to each point with kernel density estimation \cite{Logan2020a}, thereby increasing the weight of points from less sampled regions of the parameter space.

\begin{figure}
    \includegraphics[width=0.95\linewidth]{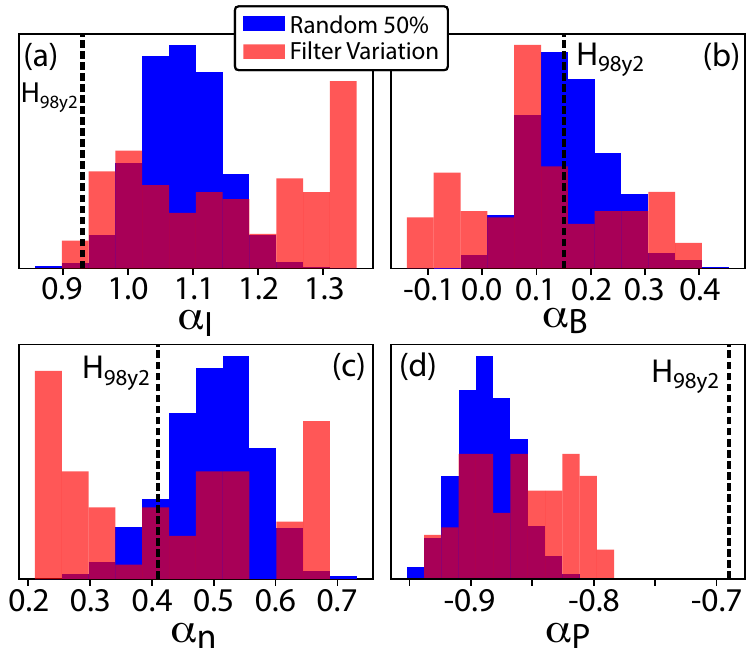}
        \vspace{-10 pt}
        \caption{Exponents of (a) \Ip{}, (b) \Bt{}, (c) \nebar{}, (d) \Ptot{} for an ensemble of \taueth{} regressions using randomized 50\% subsets of the data (blue) and different reasonable data selection criteria combinations (red). Also shown in vertical dashed lines are the exponents from the \HH{} scaling. The increased \Ip{} and \Ptot{} exponent magnitude vs. the \HH{} scaling is robust.}
    \label{fig:ensemble}
    \vspace{-15 pt}
\end{figure}

Though a regression with a good correlation coefficient can be generated, it is important to be conscious of the limitations of this technique, especially when the dataset draws from only one machine. Even when large, multi-machine datasets are used significant uncertainties can remain, as described recently in Ref. \cite{Verdoolaege2021}. With the current dataset, the main concerns are related to uncertainties originating in data selection and the potential for inconsistency between the engineering parameter scaling and scalings derived from dimensionless parameter scans \cite{Petty2008} (which are outside the scope of this letter). The importance of data selection is already illustrated by the isolation of the \fGW{} $>$ 1 points in Fig.~\ref{fig:regress}. 

Variation in the fitted confinement time scaling exponents for \Ip{}, \Bt{}, \nebar{}, and \Ptot{} are shown in Fig.~\ref{fig:ensemble}. To quantify the importance of data selection, independent regressions are created with combinations of strict, lenient, and no filters on the parameters of \Zeff{}, \fGW{}, \Pech{}, \betan{}, the temperature ratio $T_e/T_i$, and the amplitude of saturated tearing mode activity. The variation generated from the filter selections is also contrasted with the variation from randomly removing 50\% of the input datapoints. Though the exponent distributions illustrate the uncertainty in engineering regression analysis, it is clear that the NT data shows stronger dependencies on \Ip{} and \Ptot{} as compared to \HH{}. Based on the results shown in Figs.~\ref{fig:regress} and \ref{fig:ensemble}, the task of scaling NT plasmas to a reactor level of confinement should be about as challenging as the equivalent task for PT H-mode plasmas.





\paragraph{Conclusion:} 
Strongly shaped diverted NT plasmas are shown to access normalized parameters in excess of expectations for the NT regime and that compare favorably to PT H-mode plasmas. Individual access to \qnf{} $\approx$ 2.5, \betan{} $\approx$ 3.2, \fGW{} $\approx$ 1.9, and \HH{} $\approx$ 1.5 is observed, with simultaneous access to \qnf{} $<$ 3, \betan{} $>$ 3, \fGW{} $>$ 1, and \HH{} $>$ 1 demonstrated - a likely record in these simultaneous metrics across all of tokamak research. Increasing \betan{} beyond $\approx$ 2.5 and \fGW{} beyond $\approx$ 1 results in reduced \HH{} and eventually an operational limit being reached.  All results are obtained with `baseline' sawtoothing current profiles. The highest performing discharges are found at relatively low \Ip{} \& \Bt{}, motivating detailed transport analysis and projection of these discharges. The confinement scaling obtained via engineering parameter scaling is encouraging, though additional approaches will be explored to increase confidence in projection. In particular, including multiple tokamaks in confinement extrapolation studies will be essential.

This work has purposefully focused on presenting high plasma performance without model validation. Detailed modeling of the many physical mechanisms yielding the observed high performance is needed, and will be the subject of future targeted study. To summarize, improved stability is proposed to be due to the reduced size of MHD seed instabilities - namely benign sawteeth and no ELMs \cite{Nelson2023a}. Improved core density access is seen with strong density peaking set by turbulent transport \cite{Angioni2009}, with the edge density remaining within empirical limits \cite{Greenwald2002}. Finally, improved NT confinement is a well-studied area, with the largest impacts found to trapped electron mode-driven turbulent transport \cite{Camenen2007,Marinoni2019,Fontana2020}.

This work has also purposefully focused on normalized parameter access. This is because the plasma physics of interest limiting plasma current, pressure, and density scale with \qnf{}, \betan{}, and \fGW{} respectively. However, despite the good normalized parameter access observed, the absolute plasma performance did not clearly exceed existing no-ELM DIII-D plasmas \cite{PazSoldan2021}. A central reason for this is the significant reduction in plasma shape vertical elongation, which for this dataset was 1.4-1.5, as compared to the more typical 1.8-1.9 in PT, resulting in typical NT plasma volumes (14-15 m$^3$) being much lower than typical DIII-D PT plasmas (18-19 m$^3$). The reduced elongation is due to a mix of stability and operational constraints \cite{Song2021,Nelson2023}. Combined with the inherent NT penalty in \qnf{} at fixed \In{} \cite{Sauter2016}, the peak \Ip{} used in these experiments was only $\approx$ 1 MA in this dataset, compared to the routine access to \Ip{} $>$ 1.5-2.0 MA in DIII-D PT plasmas. This deficit of \Ip{} (and thus \IaB{}) precluded matching or surpassing the best absolute plasma performance without ELMs in DIII-D \cite{PazSoldan2021}. This finding motivates improvements to plasma shape control in DIII-D (and generically for NT) to increase the achievable elongation, \Ip{}, and thus \taue{} \cite{Song2021,Nelson2023}.

While this Letter has demonstrated a wide and high-performance operational space in terms of normalized core plasma parameters, significant work remains to fully appreciate the implications of these results for future NT plasmas. Evidently, interpretive modeling of the observed operational limits will be used to refine current understanding of the plasma current, pressure, and density limits and their extrapolation to future NT plasmas. Considering confinement, the presented engineering parameter regression is indeed the most basic treatment, with future work to improve projection planned via multi-machine non-dimensional scaling, transport analysis, and predictive gyrokinetic modeling of the observed results. Future work will also explore issues associated with the edge integration of these plasmas, such as operation with high radiation fraction and divertor detachment, which is outside the scope of this work.

To conclude, the observed wide operational space alongside high confinement quality increases confidence that high core performance NT plasmas can be integrated with an exhaust solution and thus develop into a viable fusion power plant. This motivates the deployment of additional resources to further advance the NT concept.


\begin{acknowledgments}
\textit{Acknowledgments:} The authors would like to especially thank additional contributors to the DIII-D NT campaign: J. Barr, W. Boyes, L. Casali, S. Ding, X. Du, D. Eldon, D. Ernst, R. Hong, F. Khabanov, R. Mattes, G. McKee, S. Mordijck, D. Shiraki, L. Schmitz, S. Stewart,  D. Truong, and A. Welsh. C. Paz-Soldan acknowledges consulting for General Atomics. Part of data analysis for this work was performed using the OMFIT integrated modeling framework \cite{Meneghini2015, Logan2018}. This material was supported by the U.S. Department of Energy, Office of Science, Office of Fusion Energy Sciences, using the DIII-D National Fusion Facility, a DOE Office of Science user facility, under Awards DE-SC0022270, DE-SC0022272, DE-SC0016154, DE-SC0014264, DE-FG02-97ER54415, and DE-FC02-04ER54698. \\
\textit{Disclaimer:} This report is prepared as an account of work sponsored by an agency of the United States Government. Neither the United States Government nor any agency thereof, nor any of their employees, makes any warranty, express or implied, or assumes any legal liability or responsibility for the accuracy, completeness, or usefulness of any information, apparatus, product, or process disclosed, or represents that its use would not infringe privately owned rights. Reference herein to any specific commercial product, process, or service by trade name, trademark, manufacturer, or otherwise, does not necessarily constitute or imply its endorsement, recommendation, or favoring by the United States Government or any agency thereof. The views and opinions of authors expressed herein do not necessarily state or reflect those of the United States Government or any agency thereof.
\end{acknowledgments}

\vspace{-10 pt}


\bibliography{library} 

\end{document}